\documentclass[%
aps,superscriptaddress,  % options for appearance
pra, % note that prl style doesn't allow appendix and removes section numbering
nofootinbib, %
reprint,%twocolumn
a4paper,% need to specify a4paper or letterpaper or watermark is messed up w/ revtex4
longbibliography
]{revtex4-1} %

\newif\ifarxiv % Comment out the next line for journal-style version
\arxivtrue

\usepackage{graphicx}
\usepackage{float}
\usepackage{amsfonts,amstext}
\usepackage{url}
\usepackage{xcolor}
\usepackage{braket}
\usepackage{soul}
\usepackage{amsthm}
\usepackage{physics}
\usepackage{hhline}
\usepackage{bbm}
\usepackage{enumerate}
\usepackage{comment}
\usepackage{colortbl}
\usepackage{amssymb}
\usepackage{comment}
\AtBeginDocument{%
    \newwrite\bibnotes
    \def\bibnotesext{Notes.bib}
    \immediate\openout\bibnotes=\jobname\bibnotesext
    \immediate\write\bibnotes{@CONTROL{REVTEX41Control}}
    \immediate\write\bibnotes{@CONTROL{%
    apsrev41Control,author="08",editor="1",pages="1",title="0",year="0"}}
     \if@filesw
     \immediate\write\@auxout{\string\citation{apsrev41Control}}%
    \fi
  }%

\usepackage[nodayofweek]{datetime}  
\usepackage{hyperref}
\definecolor{mylinkcolor}{rgb}{0,0,0.8} % set link color here as red,green,blue.
\hypersetup{unicode=true, %
  bookmarksnumbered=false,bookmarksopen=false,bookmarksopenlevel=1, %
  breaklinks=true,pdfborder={0 0 0},colorlinks=true}%
\hypersetup{%
  anchorcolor=mylinkcolor,citecolor=mylinkcolor, %
  filecolor=mylinkcolor,linkcolor=mylinkcolor, %
  menucolor=mylinkcolor,runcolor=mylinkcolor, %
  urlcolor=mylinkcolor}%

\def\q0{\underline{0}}

\renewcommand{\eqref}[1]{Eq.(\ref{#1})}

\newcommand{\be}{\begin{equation}}
\newcommand{\ee}{\end{equation}}

\begin{document}

\title{Advantages of multi-copy nonlocality distillation and its application to minimizing communication complexity} 
\author{Giorgos Eftaxias}
\email{giorgos.eftaxias@bristol.ac.uk}
\affiliation{Quantum Engineering Centre for Doctoral Training,
University of Bristol, Bristol BS8 1FD, United Kingdom}
\affiliation{Department of Mathematics, University of York, York YO10 5DD, UK}
\author{Mirjam Weilenmann}
\email{mirjam.weilenmann@oeaw.ac.at}
\affiliation{Institute for Quantum Optics and Quantum Information (IQOQI), Austrian Academy of Sciences, Boltzmanngasse 3, 1090 Vienna, Austria}
\author{Roger Colbeck}
\email{roger.colbeck@york.ac.uk}
\affiliation{Department of Mathematics, University of York, York YO10 5DD, UK}

\date{$7^{\text{th}}$ March 2023}

\begin{abstract}
Nonlocal correlations are a central feature of quantum theory, and understanding why quantum theory has a limited amount of nonlocality is a fundamental problem. Since nonlocality also has technological applications, e.g., for device-independent cryptography, it is useful to understand it as a resource and, in particular, whether and how different types of nonlocality can be interconverted. Here we focus on nonlocality distillation which involves using several copies of a nonlocal resource to generate one with more nonlocality. We introduce several distillation schemes which distil an extended part of the set of nonlocal correlations including quantum correlations. Our schemes are based on a natural set of operational procedures known as wirings that can be applied regardless of the underlying theory.  Some are sequential algorithms that repeatedly use a two-copy protocol, while others are genuine three-copy distillation protocols.  In some regions we prove that genuine three-copy protocols are strictly better than two-copy protocols. By applying our new protocols we also increase the region in which nonlocal correlations are known to give rise to trivial communication complexity.  This brings us closer to an understanding of the sets of nonlocal correlations that can be recovered from information-theoretic principles, which, in turn, enhances our understanding of what is special about quantum theory.
\end{abstract}

\maketitle

\ifarxiv\section*{Introduction}\else\noindent{\it Introduction.|}\fi A bound on the strength of correlations realisable between pairs of measurement inputs and outputs in any local theory was first shown by Bell~\cite{Bell,bell1975theory}. This bound is exceeded in quantum theory and there are even stronger correlations theoretically possible without enabling signalling~\cite{tsirelson,PR}. One way to better understand quantum theory is to consider it in light of possible alternative theories, which can be compared in terms of the correlations they can create, and the implications access to such correlations would have. For instance, it is known that theories that permit strong enough correlations have trivial communication complexity~\cite{Brassard}. Furthermore, non-local correlations have found applications in cryptography, where they form a necessary resource for device-independent quantum key distribution~\cite{Ekert91, Mayers1998,BHK,PABGMS} and randomness expansion~\cite{RogerThesis, Pironio2010, Colbeck2011}, for example. Since non-local correlations serve as resources for information processing, it is natural to ask about their interconvertability. In this work we look at non-locality distillation~\cite{Forster2009}, i.e., whether with access to several copies of some non-local resource we can generate stronger ones, which would have implications for the study of device-independent tasks in noisy regimes, for instance.

Non-locality distillation is often analysed in terms of \emph{wirings}~\cite{Forster2009,Brunner,Allcock,Hoyer,Wu2010,Chen2012,Cao2015}, which means interacting with systems by choosing inputs and receiving and processing outcomes from those systems. This has the advantage that, firstly, the distillation procedures apply to non-local quantum correlations no matter how complicated the system these have been obtained from and, secondly, these procedures are applicable beyond quantum theory. 
A general theory will prescribe various different ways to measure systems (in quantum theory, for instance, a measurement is described by a POVM). Wirings form an operationally natural sub-class that can be performed in any generalized probabilistic theory (GPT)~\cite{barrett} (including quantum theory). 

Previous work on non-locality distillation has focused on specific protocols for the distillation of 2 copies of a non-local resource (see e.g.,~\cite{Forster2009,Brunner,Allcock,Wu2010,Cao2015}).  The case of more copies remains largely open, with only few specific results~\cite{Hoyer,Chen2012}. In part, this is because analysing non-locality distillation is challenging: distillation protocols act non-linearly on the correlations and hence cannot be easily optimised. Furthermore, applying a successful 2-copy protocol twice often decreases the non-locality again (see e.g.~\cite{Brunner} for an exception). Hence, understanding 2-copy protocols provides little insight into the $n$-copy case.

In this Letter we describe a sequential adaptive algorithm that uses wirings to distil non-locality. We use this algorithm to explore the distillable region within the set of non-local correlations, and the amount of distillation possible. We demonstrate new wirings that allow distillation of correlations that cannot be distilled with any 2-copy wiring protocol. 

Our results have implications for communication complexity. In this problem, Alice with input $x$ and Bob with input $y$ want to enable Alice to compute $f(x,y):\{0,1\}^k\times\{0,1\}^m\to\{0,1\}$. We ask how much communication from Bob to Alice is required to do so. Communication complexity is said to be trivial if any such function (no matter how large $k$ and $m$) can be computed using only one bit of communication. Shared maximally non-local resources are known to make communication complexity trivial in this sense~\cite{vD}. A probabilistic notion of trivial communication complexity was introduced in~\cite{Brassard} in which for any $f$ we require the existence of $p>1/2$ such that Alice can obtain the correct value of $f(x,y)$ with probability at least $p$ for all $x$ and $y$. In this paper, when we talk about trivial communication complexity we mean it in this probabilistic sense. A larger set of shared states that render communication complexity trivial were found in Refs.~\cite{Brassard,Brunner}. Our results further enlarge this set, demonstrating advantages of wirings beyond two copies.

\bigskip

\ifarxiv\section*{Non-locality and wirings}\else\noindent{\it Non-locality and wirings.|}\fi Correlations of inputs $x,y$ and outputs $a,b$ are described by conditional probability distributions $P(ab|xy)$, and we refer to these as a \emph{box} or a \emph{behaviour}. In the context of non-locality, we usually imagine these correlations as generated by two parties, Alice and Bob, who each choose an input ($x$ and $y$ respectively) and obtain an output ($a$ and $b$ respectively). The correlations they can generate according to any theory that is consistent with special relativity have to be \emph{non-signalling}, meaning
$$
\sum_{b} P(ab|xy)=\sum_{b} P(ab|xy') \quad  \forall \ a,x,y,y',
$$
and the same holds with the roles of Alice and Bob (i.e., $a,x$ and $b,y$) exchanged. A box is called \emph{local} if it can be written
$$
P(ab|xy)=\sum_{\lambda} P(a|x\lambda)P(b|y\lambda) P(\lambda) \quad \forall \ a,b,x,y\,.
$$
In the language of Bell inequalities, there is a variable $\Lambda$ that takes the value $\lambda$ with probability $P(\lambda)$. Boxes that cannot be written in this form are \emph{non-local}.

In the case of two binary inputs and outputs, i.e., $a,b,x,y \in \{0,1 \}$, the set of all local boxes is the convex hull of $16$ local deterministic boxes $P^{\rm L}_i(ab|xy)=\delta_{a, \mu x \oplus \nu}\, \delta_{b, \sigma y \oplus \tau}$ for $\mu, \nu,\sigma,\tau \in \{0,1\}$, $i=1+\tau+2\sigma+4\nu+8\mu$, 
and the set of all non-signalling boxes is the convex hull of these local boxes and $8$ extremal non-local boxes~\cite{Cirelson93,PR} $P^{\rm NL}_i(ab|xy)=\frac{1}{2}\delta_{a \oplus b, x y \oplus \mu x \oplus \nu y \oplus \sigma}$ for $\mu, \nu, \sigma \in \{0,1\}$, $i=1+\sigma+2\nu+4\mu$.  Up to symmetry, the Clauser-Horne-Shimony-Holt (CHSH) inequality~\cite{CHSH} is the only one that restricts the set of local boxes. Non-locality can hence be quantified in terms of the CHSH value $\operatorname{CHSH}(P(ab|xy))=E_{00} + E_{01} + E_{10} - E_{11}$, with $E_{xy}=P(a=b|xy)-P(a \neq b|xy)$.

Because we work in a black-box picture, the most general operation we consider for each party is a wiring. We describe here the deterministic wirings; the most general wirings are convex combinations of these. Consider a party with access to $n$-boxes with inputs $x_j$ and outputs $a_j$ with $j=1,\ldots,n$. They ``wire'' these together to form a new box with input $x$ and output $a$. The most general deterministic wiring comprises choosing a box to make the first input to and then making a chosen input, then using the output of that box to choose the second box and the input to that second box and so on. We label the $i^{\text{th}}$ box chosen $j_i(x,a_{j_1},\ldots,a_{j_{i-1}})$
%\textcolor{green}{(I do not understand the notation-entries here)} 
and its input $x_{j_i}(x,a_{j_1},\ldots,a_{j_{i-1}})$. The final outcome is chosen depending on the overall input and all previous outcomes $a(x,a_{j_1},\ldots,a_{j_n})$. Thus, if Alice and Bob each do wirings on shares of $n$ boxes, they generate a new box $P(ab|xy)$.

\begin{figure}
\centering
\begin{minipage}[t]{0.48\columnwidth}
	\includegraphics[width=1\textwidth]{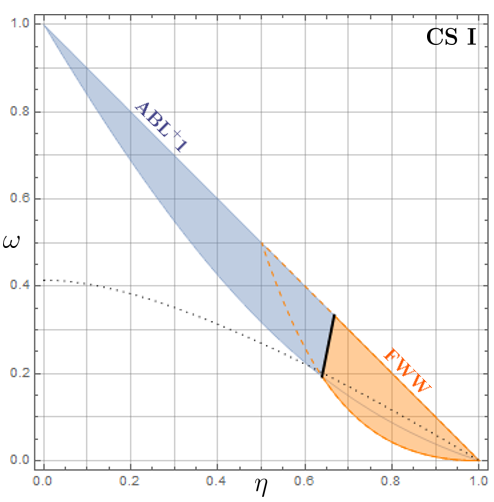}
\end{minipage}
\begin{minipage}[t]{0.05\columnwidth}
	
\end{minipage}
\begin{minipage}[t]{0.47\columnwidth}
	\includegraphics[width=1\textwidth]{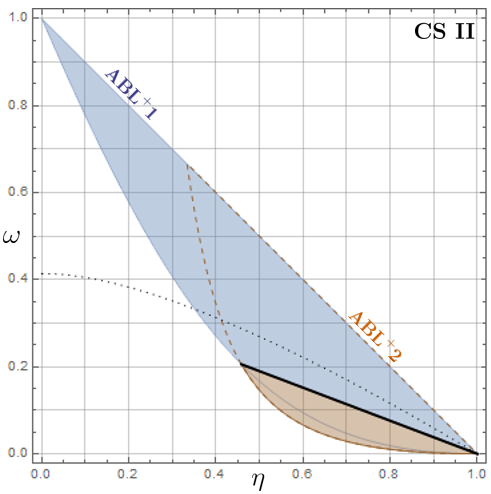}
\end{minipage}
\caption{Protocols sufficient to characterise the two-copy distillability (both the distillable region and the strongest amplification) for two CSs (cf.\ \eqref{CSequations}). The optimal two-copy protocols for CS~II are the two protocols from~\cite{Allcock} (ABL$^+$1,2), while for CS~I the protocol of~\cite{Forster2009} (FWW) is optimal in some cases. The shading indicates where the corresponding protocol is optimal, with the boundary indicated by the black line (see Appendix~\ref{sec:app1} for details of the protocols). The dotted curve indicates the boundary of the set of correlations realisable in quantum theory (computed using the conditions in~\cite{tsirelson,Masanes}).}
\label{fig:trianlges123_2copy}
\end{figure}

Our main question is then: \emph{given several copies of a non-local box, are there wirings for Alice and for Bob such that the resulting box is more non-local than the original?} In the case of two non-signalling boxes each with binary inputs and outputs, the possible wirings have been fully characterised~\cite{Short2010}. Nevertheless, even in this case, deciding whether these can result in more non-locality for a specific box is computationally intensive: there are $82$ deterministic wirings that each party can perform for each input~\cite{Short2010}, leading to a total of $82^4$ possibilities (one of the $82$ for each input of each party). To make the computation more tractable, we optimise the wirings of one party with a linear program, while iterating over $82^2$ wirings for the other (see Appendix~\ref{sec:app1} for more details). We use this linear programming technique to illustrate the regions in which distillation is possible for various 2-dimensional cross-sections (CSs) of the no-signalling polytope in Figure~\ref{fig:trianlges123_2copy}. In this work we consider three regions:
\begin{align}
\mathrm{CS~I:}&\ \omega P^{\mathrm{NL}}_1+\frac{\eta}{2}( P^{\mathrm{L}}_1+P^{\mathrm{L}}_6)+(1-\omega-\eta)P^{\mathrm{O}} \nonumber \\
\mathrm{CS~II:}&\ \omega P^{\mathrm{NL}}_1+\eta P^{\mathrm{L}}_1+(1-\omega-\eta)P^{\mathrm{O}} \label{CSequations}\\
\mathrm{CS~III:}&\ \omega P^{\mathrm{NL}}_1+\frac{\eta}{2} (P^{\mathrm{L}}_1+P^{\mathrm{L}}_9)+(1-\omega-\eta)P^{\mathrm{O}}\,,\nonumber 
\end{align}
where $P^{\mathrm{O}}=3/4P^{\mathrm{NL}}_1+1/4P^{\mathrm{NL}}_2$ is local and $\eta,\omega\geq0$ with $\eta+\omega\leq1$.

We analysed the distillability within these cross sections. Among the optimal protocols we recovered several that were previously known~\cite{Brunner2011,Allcock}. The protocols of~\cite{Allcock} (called ABL$^{+}1,2$) are sufficient to characterize the two-copy distillability in CS~II (see Figure~\ref{fig:trianlges123_2copy}), and CS~III is two-copy non-distillable. The observation that ABL$^{+}$2 achieves no distillation in CS~I shows that optimal protocols depend on the cross-section.

The above analysis is generally not useful for analysing whether repeated distillation of a box can lead to a certain CHSH-value. Applying a wiring that works for two boxes to two copies of the generated box often does not give a further increase in non-locality% but maps outside of the distillable region of that protocol
, in which case a switch of wirings is needed to distil further.  %(the protocol in~\cite{Brunner} is an exception).
While there are boxes that cannot be distilled at all with wirings (e.g.\ isotropic boxes~\cite{Beigi}), the maximum CHSH value that can be distilled using multiple copies of a specific resource box is unknown. This means that we do not know how resourceful (multiple copies of) most non-local boxes are for information processing. For instance, shared boxes render communication complexity trivial if their initial CHSH value is greater than $\operatorname{CHSH}(P(ab|xy))= 4 \sqrt{\frac{2}{3}}$~\cite{Brassard}.  The complete set of boxes that render commuication complexity trivial is unknown, although an additional region was found with the protocol of~\cite{Brunner}.

\bigskip

\begin{figure}[h]
	\centering
	\begin{minipage}[t]{0.9\columnwidth}
		\includegraphics[width=1\textwidth]{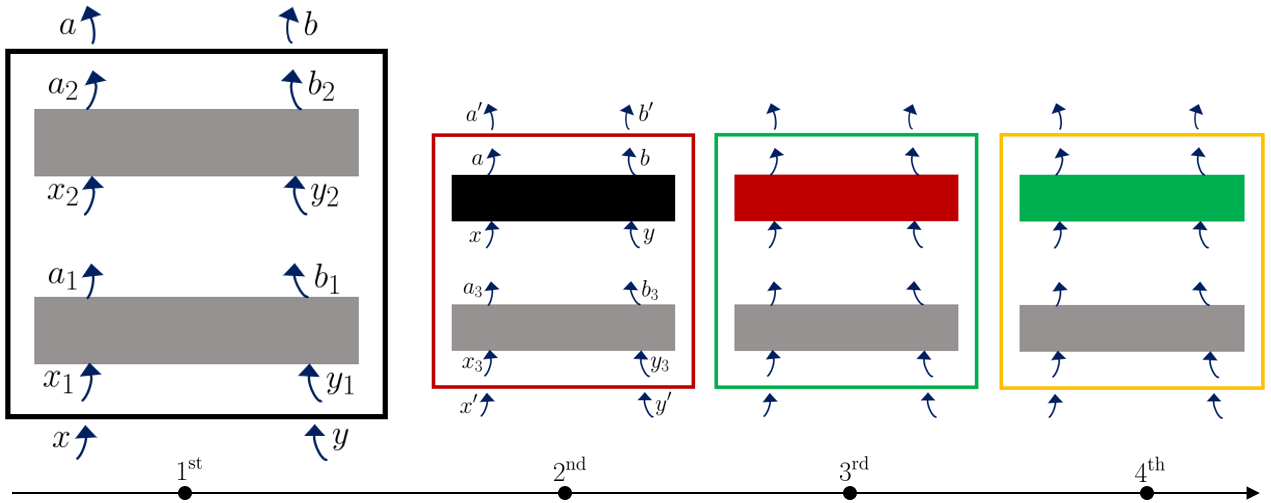} 
	\end{minipage}
	\caption{A serial architecture for combining nonlocal resources (gray) in a sequential manner. The first step on the left depicts the usual two-copy distillation scheme. Each subsequent iteration uses another copy of the original box and the previously generated one. Our sequential algorithm optimises the protocol at each round. See Appendix~\ref{sec:app2} for details.}
	\label{fig:sequential_wiring_architectures}
\end{figure}

\begin{figure}[h]
	\centering
	\includegraphics[width=0.7\columnwidth]{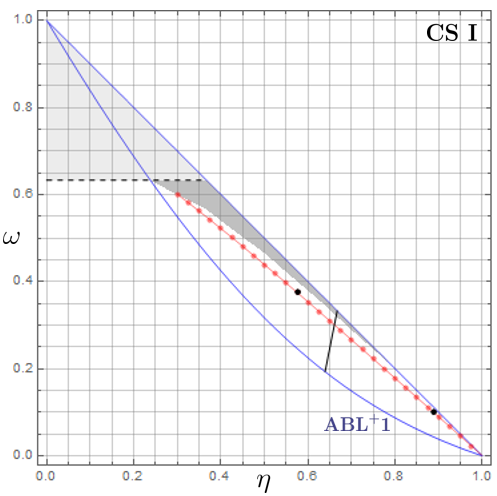}
	\caption{Region of trivial communication complexity in CS~I. The light-gray part was identified in~\cite{Brassard}. The dark-gray region includes boxes that trivialise communication complexity through (up to 4) iterations of ABL$^+$1. The red points (and everything on their right) collapse communication complexity using our serial algorithm. The black solid chord is that of Figure~\ref{fig:trianlges123_2copy} (left) and indicates a change in protocol for the red points -- see Appendix~\ref{sec:app2} for details, including analysis of the black points in the figure.}
	\label{fig:trivial2copy}
\end{figure}

\ifarxiv\section*{Sequential algorithms for non-locality distillation and reduction of communication complexity}\else\noindent{\it Sequential algorithms for non-locality distillation and reduction of communication complexity.|}\fi While a repeated application of a successful 2-copy protocol often does not increase the non-locality further, there are various ways to combine different 2-copy wirings (see Appendix~\ref{sec:app2}). Here, we focus on the specific structure illustrated in Figure~\ref{fig:sequential_wiring_architectures}. 
Our serial algorithm consists in optimising the wiring to be applied in every step, which is done in terms of a hybrid procedure of iterating over wirings and linear programming (see Appendix~\ref{sec:app2} for a detailed description of the algorithm). Applying our serial algorithm, we are able to extend the region of non-local boxes known to trivialise communication complexity, see Figure~\ref{fig:trivial2copy}. 

Our algorithm furthermore provides us with a way to systematically derive new non-locality distillation protocols for multi-copy non-locality distillation.  When performing two-steps of the serial algorithm, we find the three-copy protocol below to be successful.

In the first step, a box is created from two copies of a box $P$ with inputs (outputs) labelled $x_1, y_1$ ($a_1, b_1$) and $x_2, y_2$ ($a_2, b_2$) respectively (first step in Figure~\ref{fig:sequential_wiring_architectures}). Then this is wired to another copy of $P$, $P(a_3 b_3|x_3 y_3)$, using the functions 
\begin{align} \label{eq:addon1} 
x_1&=x=x',\ x_2=x\oplus \bar{a}_1,\ a=a_1\oplus a_2,\ x_3=x\bar{a}  \nonumber \\
y_1&=y=y',\ y_2=y b_1,\ b=b_1\oplus b_2,\ y_3=y\oplus b, \\
a'&= a\oplus a_3 ,\  b'=b\oplus b_3, \nonumber 
\end{align} 
where $\oplus$ is the logical {\sc xor} and $\bar{z}=z\oplus1$. 
This new protocol distils in CS~II a strict superset of non-local boxes compared to the previously known 3-copy distillation protocol of~\cite{Hoyer} (in contrast to CS~I where the protocol of~\cite{Hoyer} is superior).  For completeness we introduce the protocol from~\cite{Hoyer} in Appendix~\ref{sec:app3} and we refer to it as HR. The region in which the new protocol distils in CS~II is also shown in Appendix~\ref{sec:app3}.

\bigskip

\ifarxiv\section*{Genuine three-copy distillation protocols}\else\noindent\textit{Genuine three-copy distillation protocols.|}\fi When considering 3-copy distillation, the variety of possible protocols is vastly increased. In this case we can derive new protocols that outperform the previous ones in terms of the boxes for which they offer distillation. For this, we introduce a \emph{genuine three-copy distillation protocol}, which is one that cannot be reduced to a concatenation of 2-copy protocols, i.e., is not of the form of Figure~\ref{fig:sequential_wiring_architectures}.  Consider the following wiring, where $\lor$ denotes the logical {\sc or} operation:
\begin{align}
&x_1=x_2=\bar{x},\ x_3=\bar{x}a_1\lor\bar{x}a_2,\ a=a_1a_3\lor a_2a_3\lor\bar{a}_1\bar{a}_2\bar{a}_3,\nonumber\\
&y_1=y_2=y,\ y_3=yb_1\lor yb_2\lor\bar{y}\bar{b}_1\bar{b}_2, \label{eq:3copy_protocol_2} \\
&b=\bar{y}b_1b_3\lor\bar{y}b_2b_3\lor yb_1\bar{b}_3\lor yb_2\bar{b}_3\lor\bar{y}\bar{b}_1\bar{b}_2\bar{b}_3\lor y\bar{b}_1\bar{b}_2b_3. \nonumber
\end{align}

\begin{figure}[h]
	\centering
	\begin{minipage}[t]{0.48\columnwidth}
		\includegraphics[width=1\textwidth]{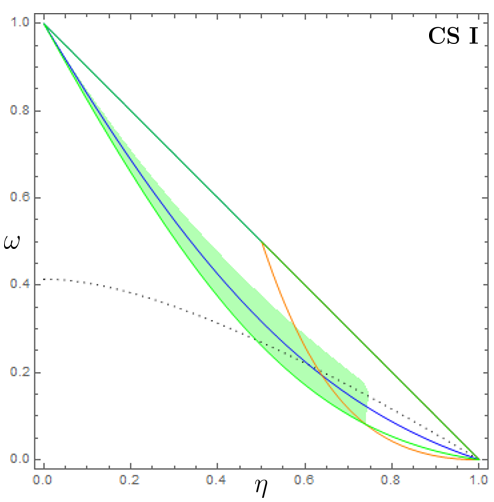}
	\end{minipage}
	\begin{minipage}[t]{0.48\columnwidth}
		\includegraphics[width=1\textwidth]{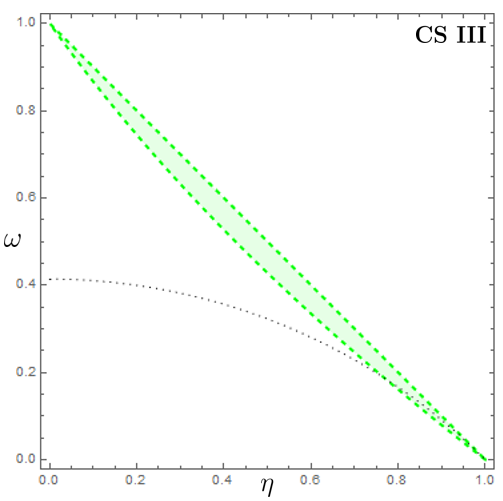}
	\end{minipage}
	\caption{Region of distillation by means of the 3-copy wiring of \eqref{eq:3copy_protocol_2} bounded by the green lines. The blue and orange lines show the region of optimal 2-copy distillation in CS~I, as in Figure~\ref{fig:trianlges123_2copy} (left). The green shaded area in CS~I depicts where our protocol leads to higher CHSH values than all previously known protocols (i.e., 2-copy and 3-copy FWW, ABL$^{+}$1, HR). In CS~III no 2-copy non-locality distillation is possible and the ability to distil is unlocked only when given access to at least 3 copies of a non-local box where use of a genuine 3-copy protocol is imperative. The dotted curve indicates the boundary of the set of quantum-realizable correlations.}
	\label{fig:3copy}
\end{figure}

We find larger regions of distillable boxes as compared to the two-copy case, see Figure~\ref{fig:3copy}.
In CS~III no 2-copy distillation is possible, while with 3 copies it is. Furthermore, the increase in the region of boxes that allow for distillation is considerably larger than that of HR (which is nearly indistinguishable from ABL$^+$1, see also Figure~\ref{fig:HRvsABL1vsStar} in the Appendix).

\begin{figure}
	\centering
	\begin{minipage}[c]{0.45\columnwidth}
		\includegraphics[width=0.9\textwidth]{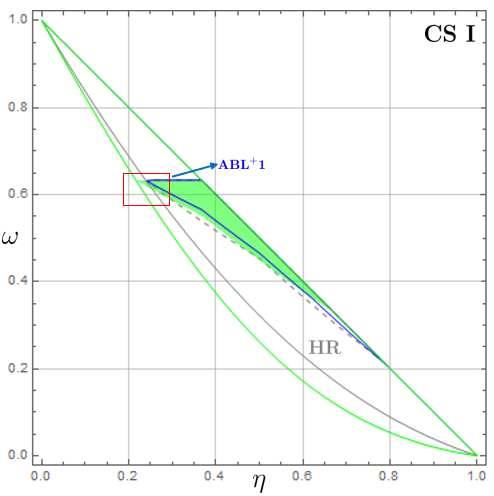}
	\end{minipage}
	\begin{minipage}[c]{0.45\columnwidth}
	\includegraphics[width=0.9\textwidth]{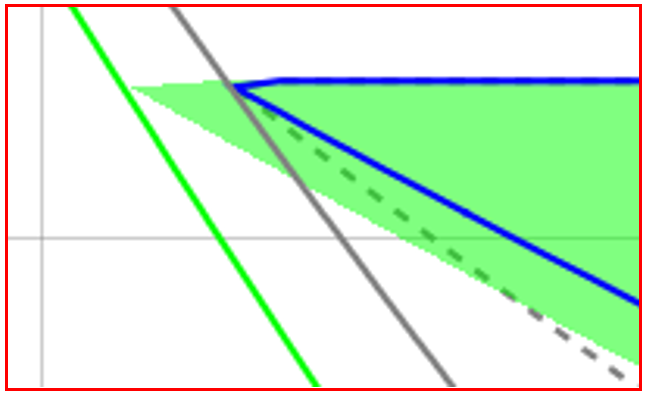}
	\end{minipage}
	\caption{Regions of trivial communication complexity with various protocols. The green region is from repeated use of our genuine 3-copy protocol of \eqref{eq:3copy_protocol}, the blue bounded region is from repeated use of ABL$^+$1 and the dashed gray bounded region is from repeated use of HR. In the magnified view (right) we see a small region where our new 3-copy protocol outperforms HR and any possible 2-copy protocol.}
	\label{fig:trivialcomm_3}
\end{figure}

Additionally we find 3-copy protocols that increase the region where communication complexity is trivial. In particular 
\begin{align}
&x_1=x_2=x,\ x_3=xa_2\lor x\bar{a}_1\lor\bar{x}\bar{a}_2a_1,\nonumber \\
&a=a_3a_2\lor a_3\bar{a}_1\lor\bar{a}_3\bar{a}_2a_1,\ y_1=y_2=y,\ y_3=yb_2\lor y \bar{b}_1,\nonumber \\
&b=b_3b_2\lor b_3\bar{b}_1\lor\bar{b}_3\bar{b}_2b_1.\label{eq:3copy_protocol} 
\end{align}
We illustrate the use of this protocol for trivialising communication complexity in Figure~\ref{fig:trivialcomm_3}. In addition, we find that in CS~I, starting from any point with $\omega>0$ on the line $\omega=1-\eta$ we can distill arbitrarily close to a PR box by repeatedly iterating this protocol (see Appendix~\ref{sec:app4}). We observe, that as compared to using 2-copy protocols (even sequentially), 3-copy protocols provide further advantages.

Additionally, all the protocols introduced here, i.e., those of~(\ref{eq:addon1}), (\ref{eq:3copy_protocol_2}) and~(\ref{eq:3copy_protocol}) work in a full dimensional subset of the space of no-signalling correlations. This space is 8 dimensional for bipartite non-signalling boxes with binary inputs and outputs. The form of our distillation protocols (and many others in the literature) implies that the difference between the initial and final CHSH value is a polynomial in the parameters of the initial box $P(ab|xy)$ and hence continuous in these parameters. Thus, for any distillable point not on the boundary of the polytope, there exists an eight-dimensional ball around it that is also distillable.

\bigskip

\ifarxiv\section*{Conclusions}\else\noindent\textit{Conclusions.|}\fi We have found a genuine 3-copy protocol that distils nonlocality for boxes in which distillation with two copies is impossible and shown that there are 3-copy protocols that outperform \emph{all} 2-copy protocols (and sequential applications thereof). 
For the latter we employed an optimization technique for 2-copy wiring protocols. Although this optimization furthers our understanding, it remains limited to cases with small numbers of inputs and outputs and there remains much more to discover about nonlocality distillation.

Whether the principle of non-trivial communication complexity~\cite{Brassard} defines a closed set of correlations~\cite{Lang} that allows for a simple characterisation and lies well between quantum and non-signalling sets is an open question of interest for the foundations of quantum theory. Indeed, finding a sensible generalised probabilistic theory that leads to a set of correlations between the non-signalling and quantum set with a simple geometric description has been a conundrum. The present work suggests that a better understanding of multi-copy non-locality distillation may give us insights into such a set, namely that of a GPT whose only restriction is imposed by the principle of non-trivial communication complexity. This would further advance the recent research program of experimentally ruling out generalised probabilistic theories due to the correlations they produce in networks~\cite{selftest1, selftest2}. 

Some of our distillation protocols work within the set of quantum correlations (see Figure~\ref{fig:3copy}). [See also~\cite{banik22} for recent work aiming to distil quantum correlations.] Being wirings, they are much simpler to perform than entanglement distillation protocols~\cite{BBPS}. It would be interesting to explore applications of these for information processing. We also remark that in recent work we have shown that non-wiring effects can be beneficial for non-locality distillation~\cite{EWC2}.

\ifarxiv\acknowledgements\else\noindent{\it Acknowledgements|}\fi GE is supported by the EPSRC grant EP/LO15730/1. MW is supported by the Lise Meitner Fellowship of the Austrian Academy of Sciences (project number M 3109-N). Some of the preliminary work for this project was performed using the Viking Cluster, a high performance computing facility at the University of York. We are grateful for computational support from the University of York High Performance Computing service, Viking, and the Research Computing team.

%\bibliography{distBIB}
%

\appendix

\onecolumngrid

\section{Optimising over all two-copy non-locality distillation protocols} \label{sec:app1}

In order to establish whether a non-local box is amenable to 2-copy non-locality distillation, it is convenient (and due to the large number of possible protocols even necessary) to find ways to search and optimise over all such protocols. This can be achieved using Linear Programming. Specifically, while iterating over the extremal wirings of one party, we can optimise the operations of the other this way.

To see how this is possible, notice that the correlations obtained from wiring two boxes $Q_{1}(a_1b_1|x_1 y_1)$, $Q_{2}(a_2b_2|x_2 y_2)$ are 
  $$P(ab|xy)=\sum_{x_i,y_i,a_i,b_i} Q_{1}(a_1b_1|x_1 y_1)Q_{2}(a_2b_2|x_2 y_2) \chi_{x}(a x_1x_2|a_1 a_2) \xi_{y}(b y_1 y_2|b_1 b_2), $$
 where  $\chi_{x}(a x_1x_2|a_1 a_2)$ and $\xi_{y}(b y_1 y_2|b_1 b_2)$ describe Alice's and Bob's wirings upon receiving input $x$ and $y$ respectively.  For a deterministic wiring, $\chi_{x}(a x_1x_2|a_1 a_2)\in\{0,1\}$ for all $a,\,a_1,\,a_2,\,x_1,\,x_2$, and the wiring $x_1=0$, $x_2=a_1$ and $a=a_1\oplus a_2$ would correspond to $\chi(a x_1x_2|a_1 a_2)=\delta_{x_1,0}\delta_{x_2,a_1}\delta_{a,a_1\oplus a_2}$, for example.
 
 A wiring on Alice's side is made up of $|x|\cdot |a|$ vectors $\chi_x(a)=(\chi_{x}(a x_1x_2|a_1 a_2))_{a_1 a_2 x_1 x_2}$. In the case of 2-inputs and 2-outputs, these are straightforward to characterise since the wirings there are exactly the allowed measurements in a generalised probabilistic theory of non-local boxes~\cite{short2006entanglement}. Specifically, to have a valid wiring in this case, it is necessary and sufficient that the output distribution on any 2-input 2-output non-signalling box returns a valid probability distribution, i.e., for any $ Q \in \{P^{\rm L}_i, P^{\rm NL}_j\}_{i,j}$
 \begin{align}
 	0 \leq  \sum_{a_{1},a_{2}, x_{1}, x_{2}} \chi_{x}(x_{1}x_{2}a|a_{1}a_{2}) Q(a_{1}a_{2}|x_{1}x_{2}) &\leq 1 \qquad \forall x,a , \label{eq:A1}\\
 	\sum_{a, a_{1},a_{2}, x_{1}, x_{2}} \chi_{x}(x_{1}x_{2}a|a_{1}a_{2}) Q(a_{1}a_{2}|x_{1}x_{2}) &= 1 \qquad \forall x.\label{eq:A2}
 \end{align}
These are linear constraints on the vectors $\chi_x(a)$.

Furthermore, $\operatorname{CHSH}(P(ab|xy))$ is a linear function of the $P(ab|xy)$, which in turn is linear in $\chi_x(a)$. Thus, we can optimise the distilled non-locality over Alice's wirings with a linear program. Although this procedure works well when Alice and Bob each hold halves of two 2-input 2-output systems, going beyond this case presents several challenges:
\begin{enumerate}
    \item With more than two systems the number of wirings on Bob's side significantly increases.
    \item Sticking with two systems but increasing the number of inputs and outputs for each system significantly increases the number of wirings.
    \item With more than two systems it is possible that the linear program optimizing over Alice's operations outputs a vector $\chi_x$ that is not a wiring.
\end{enumerate}
The presence of such \emph{non-wirings} for three systems was first noticed in~\cite{Short2010}.  In the main text we motivated the use of wirings based on maintaining validity of the results in any GPT. Allowing the non-wirings that come from such a linear program does not significantly alter the theory-independence in the sense that \eqref{eq:A1} and \eqref{eq:A2} are minimal requirements hence if no additional restrictions are placed on the theory any $\chi_x$ output by the linear program should be valid. Nevertheless it may be unnatural to allow non-wirings for Alice while restricting to wirings for Bob.  Hence one would either like to add all the non-wirings valid in \emph{any} theory to the set of Bob's possibilities, or remove non-wirings from the set of possible operations of Alice.

In the case of 2 copies of a box, in order to optimise the distilled non-locality over all wirings of Alice \emph{and} Bob, we iterate over the $82^2$ extremal wirings of Bob, as found in~\cite{short2006entanglement} and displayed in Table~\ref{table:couplers}, while optimizing Alice's wiring for each such choice with a linear program as described above.

\begin{table}
	\centering
	\begin{tabular}{|p{2.4cm}|p{2.6cm}|p{5.3cm}|p{4.0cm}|}
		\hline
		Wiring class & Number of wirings in class & Elements $\chi(a,a_1,a_2,x_1,x_2)=1$ if the following holds: (otherwise zero) &  Label of wiring for each $\mu, \nu, \sigma, \delta, \epsilon \in \{0,1\}$ \\
		\hline
		Constant  & 2 & \vspace{0.1cm} $x_1=x_2 , \hspace{0.2cm} a=\mu$ \vspace{0.1cm} &  $\mu+1$ \\
		\hline
		One-sided  &   8 & \vspace{0.1cm} $x_1=x_2=\mu , \hspace{0.2cm} a=a_{\nu +1} \oplus \sigma$ \vspace{0.1cm} &  $(4\mu+2\nu+\sigma+1)+2$\\
		\hline
		XOR-gated   & 8 & \vspace{0.1cm} $x_1=\mu , \hspace{0.2cm} x_2=\nu ,\hspace{0.2cm}  a=a_1 \oplus a_2 \oplus \sigma $ \vspace{0.1cm} &  $(4\mu+2\nu+\sigma+1)+10$ \\
		\hline
		AND-gated   & 32 & \vspace{0.1cm} $x_1=\mu , \hspace{0.2cm} x_2=\nu ,$ \vspace{0.2cm} \newline   $ a=(a_1 \oplus \sigma)(a_2 \oplus \delta) \oplus \epsilon $ \vspace{0.1cm} &  $(16\mu+8\nu+4\sigma+2\delta+\epsilon+1)+18$ \\
		\hline
		Sequential  &   32  & \vspace{0.1cm} $x_{\mu+1}=\nu , \hspace{0.2cm} x_{(\mu \oplus 1)+1}=a_{\mu+1} \oplus \sigma ,$ \vspace{0.2cm} \newline $ a=a_{(\mu \oplus 1)+1} \oplus \delta a_{\mu+1} \oplus \epsilon $ \vspace{0.1cm} &  $(16\mu+8\nu+4\sigma+2\delta+\epsilon+1)+50$\\
		\hline
	\end{tabular}
	\caption{Labelling of 2-copy wirings. To iterate over all extremal wirings for Bob, we consider all combinations of $\xi_{0}(b)$, $\xi_{1}(b)$ from the above list, i.e., $82^2$ wirings.}
	\label{table:couplers}
\end{table} 
  
Using this technique we can find whether there is a successful protocol for 2-copy non-locality distillation for any non-local box with two inputs and two outputs. In the following we illustrate this on CSs~I and~II (cf.\ \eqref{CSequations}). In both cases, the full optimisation shows that two protocols are sufficient for characterising the region of 2-copy distillation in a CS. None of the points that are not distillable with either of these protocols can be distilled with any other 2-copy wiring there. In both CSs, we can choose non-locality distillation protocols from the literature to achieve this, i.e., known protocols are among the optimal ones when considering the region of distillation. Specifically, the region of distillation of CS~I can be characterised in terms of the protocol from~\cite{Forster2009}, which we call \emph{FWW} here, as well as a protocol from~\cite{Allcock}, called \emph{ABL$^+$1} here, which are both given in the Tables~\ref{tab:sec1} and~\ref{tab:sec2}. The parameters $\omega$ and $\eta$ are chosen like in Figure~\ref{fig:trianlges123_2copy}. Since the boundary of this region can be established as those boxes $P$ for which 2-copy distillation leads to a box $P'$ such that  $\operatorname{CHSH}(P(ab|xy))= \operatorname{CHSH}(P'(ab|xy))$, this region can be characterised analytically.

\begin{table}
	\centering
	\begin{tabular}{ |p{1.6cm}|p{2.3cm}| p{5cm}|p{5cm}| } 
		\hline
		protocol name & wiring &  analytic boundary of the region of distillation ($\omega$ as a function of $\eta$) & CHSH value of the distilled box  \\ 
		\hline
		FWW~\cite{Forster2009} & $x_{1}=x_{2}=x$ \newline $y_{1}=y_{2}=y$ \newline  $a=a_{1} \oplus a_{2} $ \newline $b=b_{1} \oplus b_{2} $ &  \vspace{0.2cm} $\omega=1 - 3 \eta + 2 \sqrt{1 - 3 \eta + 3 \eta^2}$ \newline \newline  $\eta \in [1/2,1]$ \vspace{0.1cm}& \vspace{0.3cm} $\frac{1}{2}\Big[(1+\omega)^2-3\eta^2+6\eta(1+\omega)\Big]$\\
		\hline
		ABL$^{+}$1~\cite{Allcock} & $x_{1}=x$\newline $y_{1}=y$ \newline $x_{2}=x\oplus a_{1} \oplus 1$ \newline $y_{2}=yb_{1}$ \newline $a=a_{1} \oplus a_{2} \oplus 1$ \newline $b=b_{1} \oplus b_{2} \oplus 1$  & \vspace{0.2cm} $\omega=-\eta + \frac{1}{\sqrt{3}} \sqrt{3 - 4 \eta + 4 \eta^2}$ \newline \newline  $\eta \in [0,1]$ \vspace{0.1cm}& \vspace{0.3cm} $\frac{1}{4} \Big[3 \omega^2+8 \omega-\eta^2+\eta (4+6 \omega)+5\Big]$\\
		\hline
	\end{tabular}
	\caption{Optimal 2-copy distillation protocols for CS~I.}
	\label{tab:sec1}
\end{table}

\begin{table}
	\centering
	\begin{tabular}{ |p{1.6cm}|p{2.3cm}| p{5.1cm}|p{5.5cm}| } 
		\hline
protocol name & wiring & analytic boundary of the region of distillation ($\omega$ as a function of $\eta$)  & CHSH value of the distilled box \\ 
		\hline
		ABL$^{+}$2~\cite{Allcock} & $x_{1}=x_{2}=x$ \newline $y_{1}=y_{2}=y$ \newline  $a=a_{1}a_{2} $ \newline $b=b_{1}b_{2} $  & \vspace{0.2cm} $\omega=3 - 11 \eta + 2 \sqrt{3 - 18 \eta + 31 \eta^2}$ \newline \newline  $\eta \in [1/3,1]$ \vspace{0.1cm}& \vspace{0.3cm} $\frac{1}{8}\Big[\omega^2+10\omega-3\eta^2+\eta(6+22\omega)+13\Big]$\\
		\hline
		ABL$^{+}$1~\cite{Allcock} & $x_{1}=x$\newline $y_{1}=y$ \newline $x_{2}=x\oplus a_{1} \oplus 1$ \newline $y_{2}=yb_{1}$ \newline $a=a_{1} \oplus a_{2}\oplus 1$ \newline $b=b_{1} \oplus b_{2}\oplus 1$  & \vspace{0.2cm} $\omega=-\frac{4}{3}\eta + \frac{1}{3} \sqrt{9 - 18 \eta + 25 \eta^2}$ \newline \newline  $\eta \in [0,1]$ \vspace{0.1cm}& \vspace{0.3cm} $\frac{1}{4} \Big[3 \omega^2+8 \omega-3\eta^2+\eta (6+8 \omega)+5\Big]$\\
		\hline
	\end{tabular}
\caption{Optimal 2-copy distillation protocols for CS~II.}
	\label{tab:sec2}
\end{table}

The two CSs are displayed in Figure~\ref{fig:trianlges123_2copy}. The black line where the two protocols work equally well is analytically characterised as 
\begin{align}
	\omega=5\eta-3\,,\hspace{0.5cm} \frac{1}{2}(1+\frac{1}{\sqrt{13}}) \leq \eta  \leq \frac{2}{3} \label{chordI}
\end{align}
in CS~I and 
\begin{align}
	\omega = -\frac{2\sqrt{6}-3}{5}(\eta-1)\,,\hspace{0.7cm}
	\frac{1}{25} (9 + \sqrt{6}) \leq \eta \leq 1 \label{chordII}
\end{align}
in CS~II.

We remark here that previously, heuristics to simplify the optimisation over two-copy protocols have been proposed. For instance, the method in~\cite{Brito} suggests to reduce the search over $82^4$ protocols to a manageable number of only $3152$, by only considering protocols that preserve the PR-box, $P^{\rm NL}_1$. Using linear programming, as proposed here, has the advantage that it takes all distillation protocols into account.  In contrast, the heuristic from~\cite{Brito} discards various protocols, e.g., FWW and ABL$^+$2, that despite not preserving $P^{\rm NL}_1$, are useful for non-locality distillation---they are even among the optimal 2-copy distillation protocols in CS~I---so this shortcoming is pertinent. 

\section{Sequential non-locality distillation into the region of trivial communication complexity}\label{sec:app2}
\begin{figure}[h]
	\centering
	\begin{minipage}[t]{0.48\columnwidth}
		\includegraphics[width=0.9\textwidth]{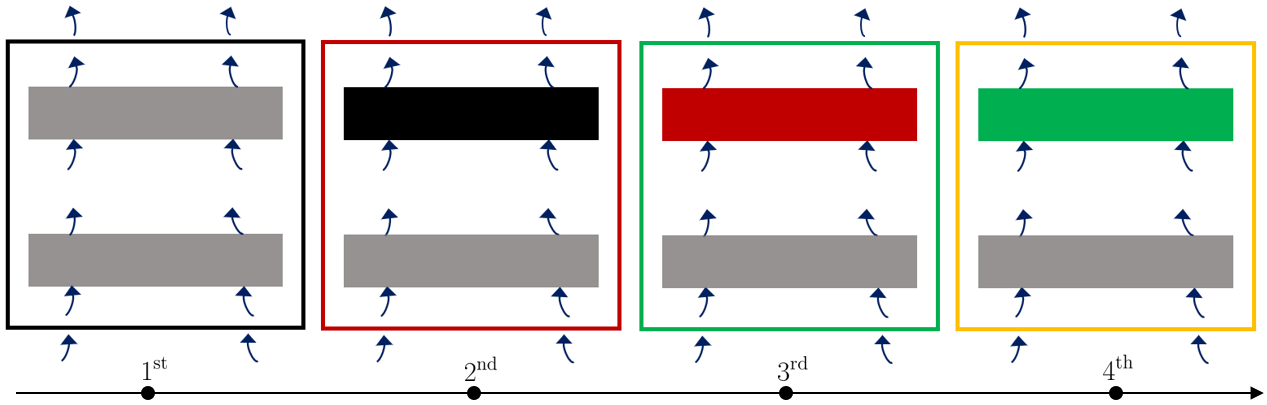}
	\end{minipage}
	\begin{minipage}[t]{0.48\columnwidth}
			\includegraphics[width=0.9\textwidth]{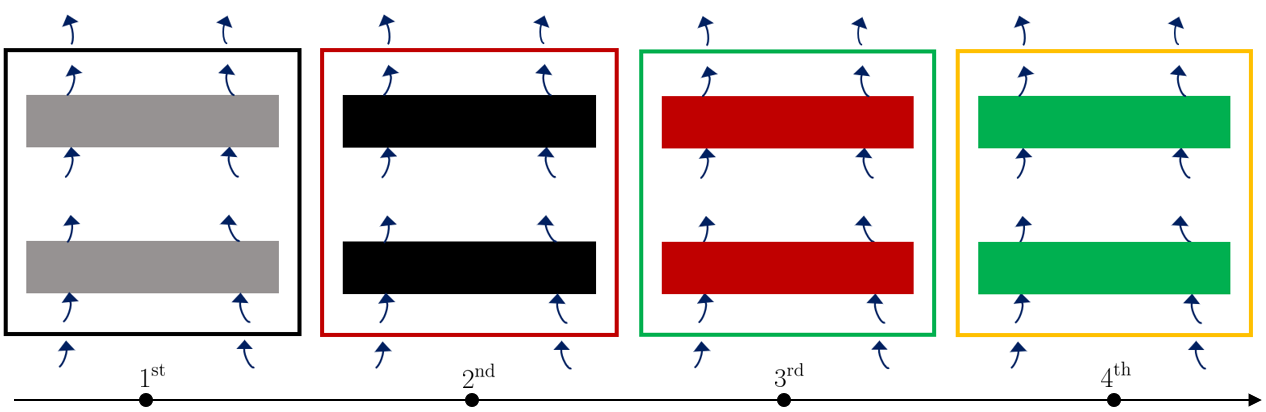}
	\end{minipage}
	\caption{Two architectures for combining an arbitrary number of resource boxes (gray) in a sequential manner. In each case, the purpose of our sequential algorithm is to find new optimal wirings in each round. Thus, the serial architecture on the left represents the \textit{serial algorithm} introduced in Figure~\ref{fig:sequential_wiring_architectures}. Similarly, the parallel architecture on the right will represent the \textit{parallel algorithm}.} 
	\label{fig:architectures2}
\end{figure}

In some situations we would like to distil non-locality up to a certain value that is useful for a specific task, e.g.\ because a particular CHSH score is needed in a device-independent scenario, or because we want to draw conclusions about the properties of those correlations, e.g.\ that they are unnatural since they imply that communication complexity is trivial. For this purpose, 2 copies of a non-local box are usually not sufficient. Since the repeated application of a fixed protocol is generally not successful in this respect either, it is natural to combine \emph{different} protocols instead. There are various ``architectures'' that such combinations can take, two of which are displayed in Figure~\ref{fig:architectures2}.

\bigskip

Analysing all of the wirings that are possible in such a multi-round procedure is computationally infeasible. We thus propose a sequential algorithm to (partially) optimise these procedures.  This algorithm (in either version of Figure~\ref{fig:architectures2}, serial or parallel or some alternatives, analysed more carefully in~\cite{GiorgosThesis}) proceeds as follows:
\begin{enumerate}[(1)]
	\item Optimise the wiring step by step using the procedure outlined in Appendix~\ref{sec:app1}. As figure of merit to be optimised we use the CHSH value here.
	\item Stop the procedure when either a certain round number is reached or when the CHSH value does not increase any further.
\end{enumerate}

\smallskip

When applying the serial algorithm to the black points from Figure~\ref{fig:trivial2copy}, choosing the serial architecture turned out to be more effective than the parallel (in terms of distilled CHSH values). The tables below compare the findings of the serial algorithm with repeated iterations of other protocols.
\begin{table}
    \centering
\begin{tabular}{ |p{0.5cm}|p{1.5cm}|p{1.5cm}|p{1.5cm}|p{1.5cm}|p{2.5cm}|p{2.5cm}|  }
 \hline
 \multicolumn{7}{|c|}{CS~I, \hspace{0.6cm} point \hspace{0.07cm} $(\eta,\omega)=(0.888, 0.1)$, \hspace{0.6cm} $\text{CHSH}_{init}=2.2$} \\
 \hline
 \multicolumn{5}{|c|}{$\text{CHSH}_{final}$ \hspace{0.03cm}, \hspace{0.2cm} after \# iterations} &
\multicolumn{2}{|c|}{Serial Algorithm \textbf{STRATEGIES}} \\
 \hline
 iter \# & two-copy ABL$^{+}$1, blindly repetitive   & two-copy FWW, blindly repetitive & two-copy \newline BS, blindly repetitive & Serial Algorithm &   Alice's wiring \hspace{0.2cm} $(\chi_{x=0} \hspace{0.1cm}, \hspace{0.1cm} \chi_{x=1}) $ & Bob's wiring \hspace{0.2cm} $(\chi_{y=0} \hspace{0.1cm}, \hspace{0.1cm} \chi_{y=1}) $\\
 \hline
 1   & 2.2815    &2.3525 & 2.2812 & 2.3525 & (12, 18)& (12, 18)\\
 \hline
 2&   2.3837 & 2.5546   & 2.3823 & 2.4681 & (12, 18) & (12, 18) \\
 \hline
 3 & \cellcolor{yellow} 2.4964  & \cellcolor{yellow} 2.7191 & \cellcolor{yellow} 2.4918 & 2.5546 & (12, 18) & (12, 18)\\
\hline
 4  & 2.5885 &  &  2.5749 & 2.6186  & (12, 18)& (12, 18) \\
 \hline
  \hline
 5 &  2.5927   & &  & 2.6729  & (12, 78) & (74, 78) \\
 \hline
 6 &     & &  & 2.7236  &(70, 82) & (12, 82)  \\
 \hline
 7 &     & &  & \cellcolor{yellow} 2.7706  & (12, 78) & (74, 78)\\ 
 \hline
 8 &     & &  &2.8143 & (70, 82) & (12, 82)  \\
 \hline
 9 &     & &  & ... & $\circlearrowleft$ & $\circlearrowleft$ \\
 \hline
 10 &     & &  & ... & $\circlearrowright$ & $\circlearrowright$ \\
 \hline
 \hline
 \hline
 36 &     & &  &3.2683  &  (70, 82) & (12, 82)   \\
  \hline
 \hline
 \hline
  41 &     & &  &3.2730  & (12, 78)  &  (74, 78)  \\
  \hline
\end{tabular}
\caption[$(\eta,\omega)_{I}=(0.888, 0.1)_I$]{Data about the lower black point of Figure~\ref{fig:trivial2copy}. The wirings are described using the labellings of the last column of Table~\ref{table:couplers}. The circular arrows denote the continued switching between the two strategies appearing on each side after the 4th iteration. The distilled CHSH values are recorded here as long as they increase. The yellow shaded entries compare final-CHSH values when each scheme has used 8 identical resource boxes. We observe that (37 copies of) the initial box trivializes communication complexity, a fact that only the serial algorithm reveals.}
    \label{(0.888, 0.1)I_strategies_displayed}
\end{table} 
\begin{figure}
    \centering
    \includegraphics[scale=0.5]{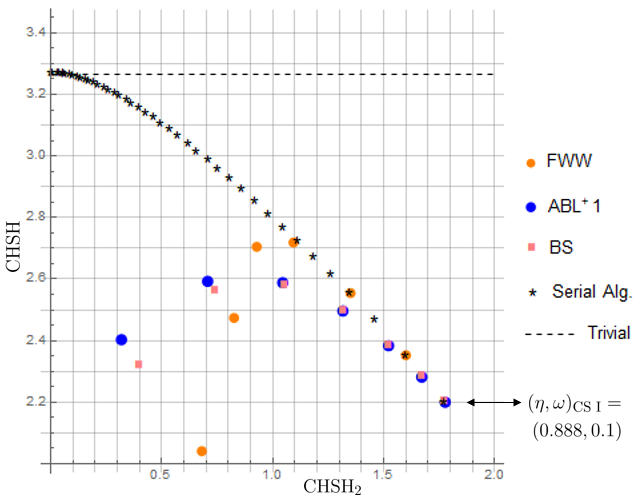}
    \caption{Visualization of the data of Table \ref{(0.888, 0.1)I_strategies_displayed} (plus some further iterations that decrease the final-CHSH value). Here, the superiority of the serial algorithm -- as opposed to the independent repetition of a fixed protocol-- makes the initial box surpass the trivial communication complexity threshold (dashed line). The horizontal axis shows CHSH$_2=E_{00}-E_{01}+E_{10}+E_{11}$.}
    \label{fig:my_label}
\end{figure}
\begin{table}
    \centering
\begin{tabular}{ |p{0.5cm}|p{1.5cm}|p{1.5cm}|p{1.5cm}|p{1.5cm}|p{2.5cm}|p{2.5cm}|  }
 \hline
 \multicolumn{7}{|c|}{CS~I, \hspace{0.6cm} point \hspace{0.07cm} $(\eta,\omega)=(0.575, 0.375)$, \hspace{0.6cm} $\text{CHSH}_{init}=2.75$} \\
 \hline
 \multicolumn{5}{|c|}{$\text{CHSH}_{final}$ \hspace{0.03cm}, \hspace{0.2cm} after \# iterations} &
\multicolumn{2}{|c|}{Serial Algorithm \textbf{STRATEGIES}} \\
 \hline
 iter \# & two-copy ABL$^{+}$1, blindly repetitive   & two-copy FWW, blindly repetitive & two-copy \newline BS, blindly repetitive & Serial Algorithm &   Alice's wiring \hspace{0.2cm} $(\chi_{x=0} \hspace{0.1cm}, \hspace{0.1cm} \chi_{x=1}) $ & Bob's wiring \hspace{0.2cm} $(\chi_{y=0} \hspace{0.1cm}, \hspace{0.1cm} \chi_{y=1}) $\\
 \hline
 1   & 2.9212    &2.8212 & 2.9162 & 2.9212 & (12, 78) & (74, 78)\\
 \hline
 2&  \cellcolor{yellow} 3.0294 & \cellcolor{yellow}   & \cellcolor{yellow} 3.0096 & 3.0452 & (70, 82) & (12, 82) \\
 \hline
 3 &   & & & \cellcolor{yellow} 3.1327 & $\circlearrowleft$ & $\circlearrowleft$\\
\hline
 4  & &  &  & 3.1930  & $\circlearrowright$ & $\circlearrowright$ \\
  \hline
 5 &   & &  & 3.2324  & $\circlearrowleft$ & $\circlearrowleft$\\
 \hline
 6 &     & &  & 3.2562  & $\circlearrowright$ & $\circlearrowright$ \\
 \hline
 7 &     & &  &  3.2683  & (12, 78) & (74, 78)\\ 
 \hline
 8 &     & &  &3.2718 & (70, 82) & (12, 82)  \\
 \hline
\end{tabular}
\caption[$(\eta,\omega)_{I}=(0.575,0.375)_I$]{Data for the higher black point of Figure~\ref{fig:trivial2copy}. The wirings are described using the labellings of the last column of Table~\ref{table:couplers}. The circular arrows denote the continued switching between the two strategies appearing on each side after the 4th iteration. The distilled CHSH values are recorded here as long as they increase. The yellow shaded entries compare final CHSH values when each scheme has used 4 identical resource boxes. We observe that (8 copies of) the initial box  trivializes communication complexity, and again, this is only revealed using the serial algorithm.}
    \label{(0.575,0.375)I_strategies_displayed}
\end{table}
We can furthermore compare the different types of procedure. While we find that in CSs~I and~II, the serial procedure is more successful with respect to the increase in non-locality that is achieved, we have found other CSs where the parallel is favourable. For more details and the analysis of further types of procedures we refer to~\cite{GiorgosThesis}.

Notice also that, after a few iterations, we recover the same iteration of wiring strategies for each party in the two tables. This procedure corresponds to essentially exchanging  the roles of the two players between iterations (and some bit-flips):
\begin{align*}
    &{\bf ODD \ iterations:} \ \ \ x_2=x, \ x_1=xa_2, \ a=a_1 \oplus {a_2} \oplus 1, \ y_2=y, \ y_1=y \oplus {b_2} \oplus 1, \ b=b_1 \oplus b_2 \oplus 1 \\
    &{\bf EVEN \ iterations:} \ x_2=x, \ x_1=x \oplus a_2, \ a=a_1 \oplus {a_2} \oplus 1, \ y_2=y, \ y_1=y ( {b_2} \oplus 1) ,\ b=b_1 \oplus b_2 \oplus 1 .
\end{align*}

%\vspace{-5.5mm}
\section{3-copy distillation in the literature} \label{sec:app3}
So far, the 3-copy non-locality distillation protocol  that was so far known in the literature was introduced in~\cite{Hoyer}. This is specified by the following functions that make up the protocol HR: 
%\begin{table}[H]
 {\begin{center}
     \begin{tabular}{ |p{5cm}|p{5cm}| } 
\hline
     Alice's side & Bob's side\\
 \hline
 $x_{1}=x$ & $y_{1}=y$ \\
 $x_{2}=x\oplus a_{1}$ & $y_{2}=y\overline{b}_{1}$ \\  $x_{3}=a_{2}\overline{a}_{1}\oplus x(a_{1}\oplus a_{2} \oplus a_{1}a_{2})$ &    $y_{3} =\overline{b}_{1}\oplus b_{2}\overline{b}_{1} \oplus y(\overline{b}_{2}\oplus b_{1}b_{2})$  \\ 
 $a=a_{1} \oplus a_{2}\oplus a_{3} $ &  $b =b_{1} \oplus b_{2}\oplus b_{3} $\\
 \hline
\end{tabular}
\end{center}
}

\begin{figure}
    \centering
    \includegraphics[scale=0.48]{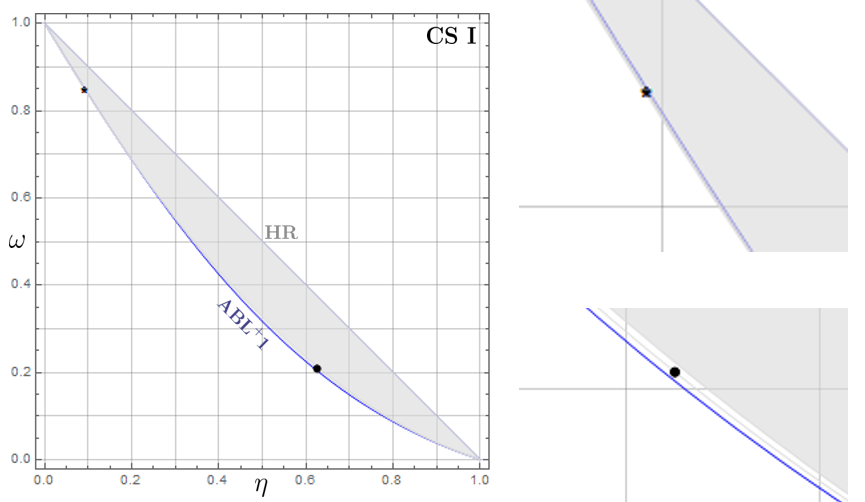}
    \caption{The blue (gray) boundary includes the boxes that are distillable by the ABL$^{+}$1 (HR). The gray region depicts the set where HR achieves higher distilled CHSH-values than ABL$^{+}$1. The bullet point corresponds to a box that is distillable by ABL$^{+}$1 but not by HR. Interestingly, the star (coordinates $(\eta,\omega)= (\frac{3}{32},\frac{1}{32}(2\sqrt{227}-3))$)  corresponds to a box that is not distillable by ABL$^{+}$1 (so, not distillable by any two-copy protocol) but it can be distilled by HR. }
    \label{fig:HRvsABL1vsStar}
\end{figure}

In some parts of CS~I, this protocol outperforms the 2-copy distillation protocol ABL$^+$1 (around the point indicated with the star in Figure~\ref{fig:HRvsABL1vsStar}). At the point indicated with the star, HR can distill non-locality while \emph{no} 2-copy protocol can (thus, HR is also a \emph{genuine} 3-copy protocol). However, the region around the starred point where this is possible is extremely small (see Figure~\ref{fig:HRvsABL1vsStar}). This is different for our genuine 3-copy protocols (Equations~(\ref{eq:3copy_protocol_2}) and (\ref{eq:3copy_protocol})), for which this increase is considerable. Furthermore, we checked that HR, despite being a genuine 3-copy protocol, distills nothing in CS~III, unlike our genuine 3-copy protocol that unlocks distillation there (Figure~\ref{fig:3copy}).

\begin{figure}
	\centering
	\includegraphics[scale=0.45]{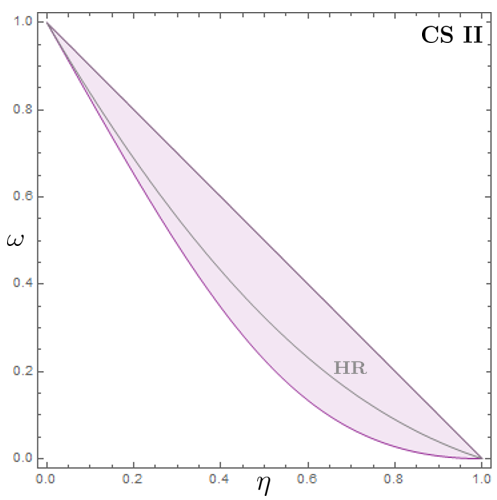}
	\caption{Comparison of our new protocol (Equation~(\ref{eq:addon1})), displayed with the purple boundary, to the 3-copy protocol HR from~\cite{Hoyer}, displayed in gray for CS~II. The protocol of Equation~(\ref{eq:addon1}) distils a strict superset of the boxes that HR distils and the non-locality increase at each point is also stronger than that of HR.}
	\label{fig:add-on}
\end{figure}

\section{Further properties of the novel OR-gated protocols}\label{sec:app4}
\begin{figure}[h]
    \centering
    \includegraphics[scale=0.7]{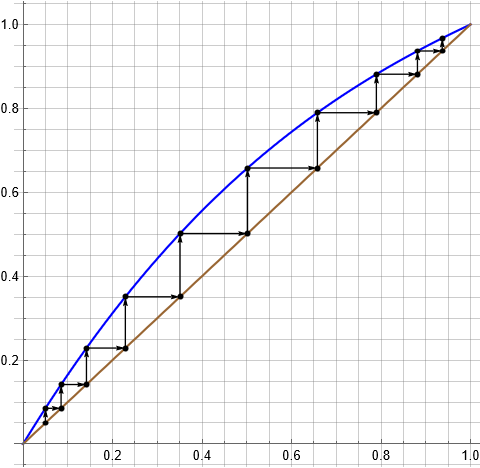}
    \caption{The blue curve depicts the function $f_1(\omega)=\frac{1}{4}\omega^2(7-4\omega+\omega^2)$ while the brown the $f_2(\omega)=\omega$, $\omega \in [0,1]$. The black arrows lying in between represent all the steps from $n=1$ to $n=10$ of the recurrence relation (\ref{recurrence}) for the case $\omega_0=0.05$.} \label{stepsfigure}
\end{figure}

In this section we present some extra features of the protocols introduced in Equations (\ref{eq:3copy_protocol_2}) and (\ref{eq:3copy_protocol}).
\begin{table}
	\centering
	\begin{tabular}{ |p{2cm}|p{9cm}| } 
		\hline
Cross Section & CHSH value of the distilled box \\ 
		\hline
		\vspace{0.3cm} \hspace{0.5cm} \Large I & \vspace{0.3cm} {\small $ \frac{1}{16}\Big[\omega^3-5\eta^3+9\omega^2+31\omega+\eta^2(5+7\omega)+\eta(9+22\omega+5\omega^2)+23\Big]$} \vspace{0.3cm} \\
		\hline
		\vspace{0.3cm} \hspace{0.5cm} \Large III & \vspace{0.3cm} {\small $ \frac{1}{16}\Big[\omega^3+7\eta^3+9\omega^2+31\omega+\eta^2(5+19\omega)+\eta(-3+18\omega+13\omega^2)+23\Big]$} \vspace{0.3cm} \\
		\hline
	\end{tabular}
\caption{Final CHSH function after one iteration of the protocol of Equation (\ref{eq:3copy_protocol_2}), for the two cross sections of Figure~\ref{fig:3copy}.}
	\label{tab:sec2p}
\end{table}
The protocol of Equation~(\ref{eq:3copy_protocol}) preserves the line (one dimensional convex combination)
\begin{align*}
    \omega P^{\mathrm{NL}}_1+(1-\omega)\frac{P^{\mathrm{L}}_1+P^{\mathrm{L}}_6}{2},
\end{align*}
which is that subset of CS~I corresponding to $\omega=1-\eta$. This means that an operation of the protocol maps any box belonging to that line, back to that line. Each iteration $n$, $n\ge 1$, of the protocol, updates the coordinate $\omega$ according to the recurrence relation
\begin{align}
    \omega_n=\frac{1}{4}\omega_{n-1}(7-4\omega_{n-1}+\omega^2_{n-1}) \hspace{0.5cm}, \hspace{0.5cm} \omega_0=\omega. \label{recurrence}
\end{align}
A plot showing the sequence of steps starting at $\omega_0=0.05$ is shown in Figure~\ref{stepsfigure}. From the shape of the curves it is clear that for any initial $\omega\in(0,1)$ repeated iterations allow us to generate a final box arbitrarily close to a PR box.

\end{document}